\documentclass{aa}

\usepackage{amssymb}
\usepackage{amsmath}
\usepackage{graphicx}
\usepackage[varg]{txfonts}
\usepackage{natbib}
\usepackage{color}

\newcommand{\gcc}{\ensuremath{\text{g} \, \text{cm}^{-3}}}
\newcommand{\kms}{\ensuremath{\text{km} \, \text{s}^{-1}}}
\newcommand{\rads}{\ensuremath{\text{rad} \, \text{s}^{-1}}}
\newcommand{\msol}{\ensuremath{M_\odot}}
\newcommand{\mch}{\ensuremath{M_\text{Ch}}}

\newcommand{\ncarb}{\ensuremath{^{12}\text{C}}}

\newcommand{\nox}{\ensuremath{^{16}\text{O}}}
\newcommand{\nsi}{\ensuremath{^{28}\text{Si}}}

\newcommand{\nni}{\ensuremath{^{56}\text{Ni}}}

\begin{document}

\title{Thermonuclear explosions of rapidly differentially rotating
  white dwarfs: Candidates for superluminous Type~Ia supernovae?}
\titlerunning{Thermonuclear explosions of rapidly differentially
  rotating WDs: Candidates for superluminous SNe~Ia?}

\author{M.~Fink\inst{1,4}
  \and   M.~Kromer\inst{2,3}
  \and   W.~Hillebrandt\inst{4}
  \and   F.~K.~R\"opke\inst{2,3}
  \and   R.~Pakmor\inst{3}
  \and   I.~R.~Seitenzahl\inst{5,6}
  \and   S.~A.~Sim\inst{7}}
\institute{%
  Institut f\"ur Theoretische Physik und Astrophysik, Universit\"at
  W\"urzburg, Emil-Fischer-Stra{\ss}e 31, 97074 W\"urzburg,
  Germany
  \and
  Zentrum f\"ur Astronomie der Universit\"at Heidelberg, Institut f\"ur
  Theoretische Astrophysik, Philosophenweg 12, 69120, Heidelberg, Germany
  \and
  Heidelberger Institut f\"ur Theoretische Studien,
  Schloss-Wolfsbrunnenweg 35, 69118 Heidelberg, Germany
  \and
  Max-Planck-Institut f\"ur Astrophysik,
  Karl-Schwarzschild-Stra{\ss}e 1, 85748 Garching, Germany
  \and
  School of Physical, Environmental and Mathematical Sciences, 
  University of New South Wales, Australian Defence Force Academy, 
  Canberra, ACT 2600, Australia
  \and 
  Research School of Astronomy and Astrophysics,
  Australian National University, Canberra, ACT 2611, Australia
  \and
  Astrophysics Research Centre, School of Mathematics and Physics, 
  Queen's University Belfast, Belfast BT7 1NN, Northern Ireland, UK
}
  
\date{Received 22 May 2018 / Accepted 17 July 2018}

\abstract{The observed sub-class of ``superluminous'' Type Ia supernovae
  lacks a convincing theoretical explanation. If the emission of such
  objects were powered exclusively by radioactive decay of
  \nni\ formed in the explosion, a progenitor mass close to or even
  above the Chandrasekhar limit for a non-rotating white dwarf star
  would be required. Masses significantly exceeding this limit can be
  supported by differential rotation. We, therefore, explore explosions
  and predict observables for various scenarios resulting from
  differentially rotating carbon--oxygen white dwarfs close to their
  respective limit of stability. Specifically, we have investigated a prompt
  detonation model, detonations following an initial deflagration
  phase (``delayed detonation'' models), and a pure deflagration
  model. In postprocessing steps, we performed nucleosynthesis and
  three-dimensional radiative transfer calculations, that allow us,
  for the first time, to consistently derive synthetic observables
  from our models. We find that all explosion scenarios involving detonations
  produce very bright events. The observables predicted for them,
  however, are inconsistent with any known subclass of Type Ia
  supernovae. Pure deflagrations resemble 2002cx-like supernovae and
  may contribute to this class. We discuss implications of our
  findings for the explosion mechanism and for the existence of differentially rotating
  white dwarfs as supernova progenitors.}


\maketitle

\section{Introduction}
\label{sec:intro}

Comprehensive observational surveys of SNe~Ia
have revealed that, despite their relative homogeneity, several
distinct subclasses exist, and the observed heterogeneity may call for
different progenitors and/or explosion mechanisms. Particularly
puzzling are supernovae that are very luminous and have decline rates
that put them well above the Phillips relation of ``normal'' SNe~Ia by
almost one magnitude in the B-band, the prototypical examples being SN
2006gz \citep{hicken2007a,maeda2009a} and SN 2009dc
\citep{yamanaka2009a,tanaka2010a,silverman2011a,taubenberger2011a}.
By now, only a few objects have been discovered that belong to this class
\citep{taubenberger2017a}. In addition to their high luminosity, 
two to three times that of normal SNe~Ia, their lightcurves
have a long rise time ($> 23$ days) and decline slowly ($\Delta m_{15}(B)
\sim 0.8$\,mag). Moreover, they are characterized by low ejecta velocities, 
occasionally being even lower than those of normal SNe~Ia, and
prominent C~\textsc{ii} absorption features, while all other
early-time spectral properties are not unusual.

These properties are not easy to reconcile within the framework of
standard explosion models. If the luminosity at peak resulted
exclusively from the decay of \nni, the Ni-mass produced in the event
would be very close to \citep{howell2006a} or even exceed the
canonical Chandrasekhar mass of non-rotating WDs. For
example, for SN~2009dc \citet{hachinger2012a} and
\citet{taubenberger2013a} find about $1.5$ to $1.8\,\msol$ of
\nni. This led to the suspicion that their progenitors might be
super-Chandrasekhar mass WDs \citep{howell2006a}, and the name
``super-Chandras'' was coined for the class. Such an interpretation sounds
reasonable because it is known that differentially rotating WDs, in
principle, can have masses up to (or even beyond) 2.5\,\msol\ (e.g.,
\citealt{durisen1975a}). The first simulations of thermonuclear
explosions of such objects \citep{steinmetz1992a} showed that a huge
amount of \nni\ is produced when they are burned in a supersonic
detonation.

However, if this is true, one has to explain why their ejecta
velocities are so low. Fusing 1.5\,\msol\ of carbon and oxygen to
iron-group and intermediate-mass elements releases roughly
$2.5\times10^{51}$\,erg of nuclear binding energy and, thus, the
kinetic energy of the ejecta should be around $1.5\times10^{51}$\,erg
(or more), inconsistent with the observed low velocities. Also,
\citet{hachinger2012a} have shown by means of ``abundance tomography''
that, at least for SN~2009dc, the amount of burned material at
high-velocity predicted by sufficiently luminous explosion models is
inconsistent with their spectra. As a way out, very asymmetric
explosions with lopsided \nni\ distributions of otherwise ``normal''
WDs were suggested \citep{hillebrandt2007a}, but this explanation
needs fine tuning. Ejecta--CSM interactions with a dense
carbon--oxygen envelope were proposed as an alternative way to explain
the high-luminosity of SN~2009dc \citep[e.g.][]{hachinger2012a,
  taubenberger2013a, noebauer2016a}.

Here, we have revisited explosions in rapidly-rotating WDs, but in
addition to the work of \citet{pfannes2010b, pfannes2010a} who
computed pure deflagration and pure detonation models only, we also
investigate deflagrations followed by a spontaneous
deflagration-to-detonation transition (DDT), the more popular scenario
for explosions of Chandrasekhar-mass WDs. As in \citet{pfannes2010b,
  pfannes2010a}, we have constructed initial models based on the work
of \citet{yoon2004a, yoon2005b}, with minor modifications due to a
slightly different equation of state (see
Sect.~\ref{sec:msetup_inimod}). An improved description of the
detonation front (see Sect.~\ref{sec:nmethods_hydro}) constitutes a
major difference to the modeling approach of \citet{pfannes2010a,
  pfannes2010b}. We also relaxed their assumption of rotational
symmetry in the gravitational potential. For all our models we ran a
consistent pipeline in which the nucleosynthesis is determined from
postprocessing the results of the hydrodynamical explosion simulations
and synthetic observables are obtained from radiative transfer
calculations. This enables, for the first time, a direct comparison of
the rapidly-rotating model predictions with observations of
superluminous SNe~Ia.

This article is structured as follows. In Sect.~\ref{sec:msetup} we
describe the initial WD models used in our work and explain how the
explosions were initiated and how the transition from the deflagration
phase to a detonation was triggered for the delayed detonation models.
In Sect.~\ref{sec:nmethods} we give a brief account of the numerical
methods used for solving the reactive Euler equations, the
nucleosynthesis post-processing step, and the radiative transfer.  We
present the results of our explosion simulations in
Sect.~\ref{sec:sresults}.  the synthetic lightcurves and spectra
obtained from these in Sect.~\ref{sec:synth_obs}. A discussion and
conclusions follow in Sects.~\ref{sec:discussion}
and~\ref{sec:conclusions}.

\section{Model setup}
\label{sec:msetup}

\subsection{Rotating initial WDs}
\label{sec:msetup_inimod}

In the single-degenerate scenario of SNe~Ia, an already rather massive
C+O WD accretes matter from a non-degenerate companion, presumably a
main sequence or subgiant star, approaches the Chandrasekhar mass
limit, $M \simeq 1.4~\msol$, and explodes. However, together with the matter,
the WD also accretes angular momentum. If this additional angular
momentum is not lost from the WD, for instance by a wind or magnetic
braking, the WD will be spun up and, due to the centrifugal force,
the critical mass for the final contraction and explosion will
increase.

Whether this angular momentum has an important impact on the
final outcome depends on the way it is redistributed in the WD\@. If
there would be no angular momentum transport from the accreted layer
into the WD a Keplerian disk would form and the accretion rate would
be low. On the other hand, hydrodynamic instabilities, such as the
(dynamical and secular) shear instability, will lead to angular
momentum transport into the WD, as studied by \citet{yoon2004a}. They
found that rapid differential rotation profiles might be established,
with a maximum angular velocity somewhere at intermediate stellar mass
shells, and a stable configuration would require that the angular
momentum is constant on cylinders around the axis of rotation.
Although their stellar evolution calculations are 1D and much
of the relevant physics had to be put-in in parametrized form,
WD masses of up to about $2\,\msol$ seem to be possible with
rather well motivated assumptions.

As in \citet{pfannes2010b,pfannes2010a} we have constructed initial models
based on the work of \citet{yoon2004a} and \citet{yoon2004b}
and have used them to run pure detonation, delayed detonation, and
pure deflagration explosion models. We have constructed the
rapidly-rotating C+O WDs in hydrostatic equilibrium, and they resemble
a subset of those presented in \citet{pfannes2010a}
(see Table~\ref{tab:msetup_inimod_param} for the model
parameters).\footnote{There are slight differences in the
  initial models of this study with respect to those in
  \citet{pfannes2010a}  due to minor differences in the equation
  of state used here.}
\begin{table}
  \centering
  \caption{Initial model parameters.}
  \label{tab:msetup_inimod_param}
  \begin{tabular}{rccc}
    \hline
    \hline
    model & AWD$1$ & AWD$4$ & AWD$3$ \\
    \hline\\[-1.7ex]
    $\rho_\text{c}~[10^9~{\gcc}]$ & $2.0$ & $2.0$ & $2.0$ \\
    $\Omega_\text{c}~[\rads]$ & $1.659$ & $4.663$ & $4.081$ \\
    $\Omega_\text{peak}~[\rads]$ & $4.473$ & $5.239$ & $5.299$ \\
    $M^\text{a}~[{\msol}]$ & $1.622$ & $1.775$ & $2.004$ \\
    $r_\text{equator}/r_\text{pole}$ & $1.629$ & $1.796$ & $2.183$ \\
    $r_\text{equator}~[10^8~\text{cm}]$ & $3.23$ & $3.32$ & $4.02$ \\
    $r_\text{pole}~[10^8~\text{cm}]$ & $1.98$ & $1.85$ & $1.84$ \\
    ${E_\text{grav}}^\text{a}~[10^{50}~\text{erg}]$ & $-36.2$ & $-40.9$ & $-44.9$ \\
    ${E_\text{int}}^\text{a}~[10^{50}~\text{erg}]$ & $27.2$ & $29.0$ & $29.5$ \\
    ${E_\text{rot}}^\text{a}~[10^{50}~\text{erg}]$ & $1.77$ & $3.01$ & $4.58$ \\
    ${E_\text{bind}}^\text{a}~[10^{50}~\text{erg}]$ & $-7.22$ & $-8.90$ & $-10.8$ \\
    $\beta^\text{a}~[\%]$ & $4.867$ & $7.342$ & $10.16$\\
    $J~[10^{50}~\text{g} \, \text{cm}^2 \, \text{s}^{-1}]$
    & $0.9110$& $1.352$ & $2.211$ \\
    \hline
  \end{tabular}
  \begin{list}{}{}
  \item[$^{\text{a}}$] These quantities have been determined after
    mapping the initial models on the hydrodynamic grid.
  \end{list}
\end{table}
To this end we applied the method of \citet{eriguchi1985a} that solves
the equations of hydrostatic equilibrium in integral form for a given
rotation law, central density and axis ratio $r_\text{equator} /
r_\text{pole}$.  The models AWD$1$, AWD$4$, and AWD$3$ (``AWD'' stands
for accreting white dwarf) span the range of total masses $M$
($1.6$--$2.0\,\msol$) and angular momentum $J$ ($0.9$--$2.2 \times
10^{50}\,\text{g} \, \text{cm}^2 \, \text{s}^{-1}$) expected for
rapidly rotating WDs formed through accretion in binary systems
(masses above $M \sim 2.0\,\msol$ are unlikely due to the limited mass
budget in single degenerate progenitor systems;
\citealt{langer2000a}).

\begin{figure}
  \centering
  \includegraphics[width=\columnwidth]{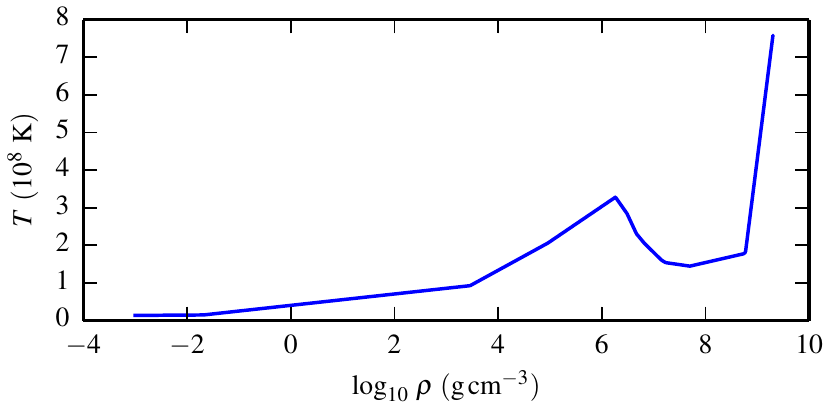}
  \caption{Initial temperature profile of the models as a function of
    density. The steep increase at high density is due to hydrostatic
    carbon burning and the local maximum at $\rho \simeq 10^6\,\gcc$
    is caused by accretion-induced heating.}
  \label{fig:msetup_temp_profile}
\end{figure}

In the initial models, we assume equal amounts (by mass) of carbon and
oxygen and no enhancement of the chemical composition due to the
metallicity of the progenitors. A temperature profile as suggested
by \citet{yoon2004a} for an accreting WD was chosen similar to
\citet{pfannes2010b,pfannes2010a} and \citet{pfannes2006a}
(see Fig.~\ref{fig:msetup_temp_profile}).
The high temperatures seen in Fig.~\ref{fig:msetup_temp_profile} arise
from the ``simmering'' C-burning phase prior to flame formation.
Convection causes an approximately adiabatic temperature gradient in
the outer layers \citep{piro2008b}. The local maximum in the outer
part of the degenerate C+O core is caused by accretion-induced heating
\citep[see][]{yoon2004a}. In all models, a central density of
$\rho_\text{c} = 2 \times 10^9\,\gcc$ was chosen.

\begin{figure}
  \centering
  \includegraphics[width=\columnwidth]{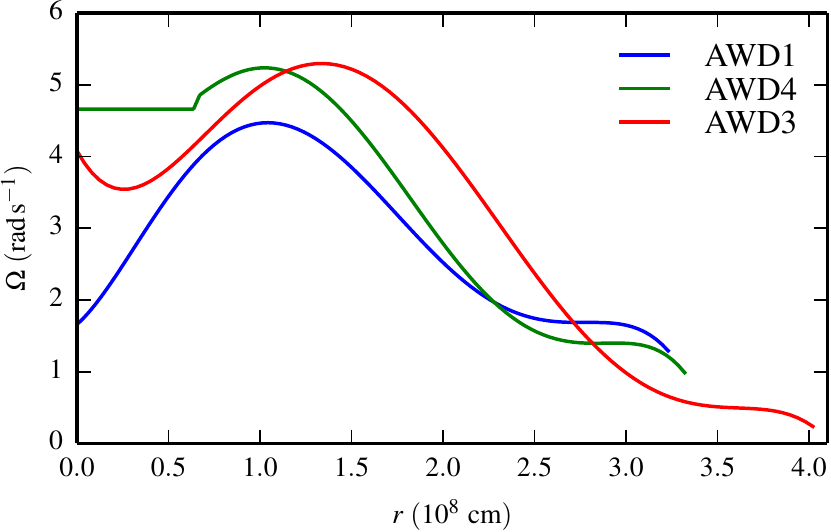}\\[0.7em]
  (a) AWD$1$, $M = 1.62~\msol$\\
  \includegraphics[width=6.8cm]{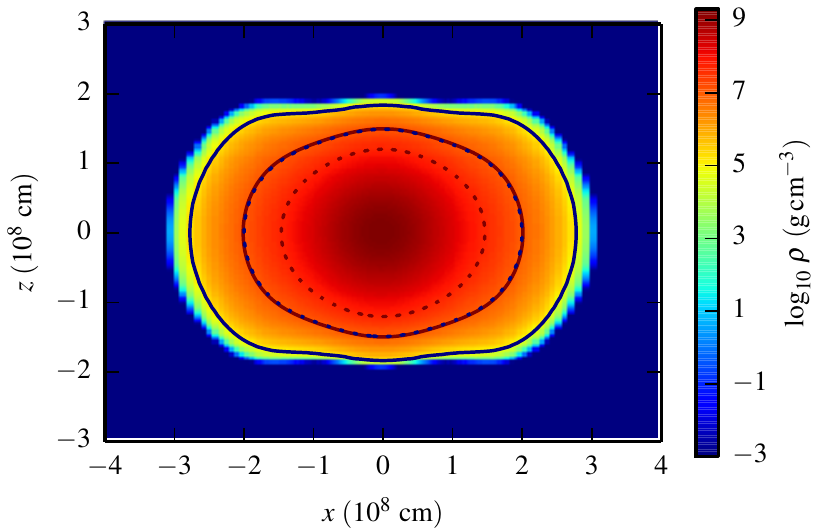}\\[0.7em]
  (b) AWD$4$, $M = 1.76~\msol$\\
  \includegraphics[width=6.8cm]{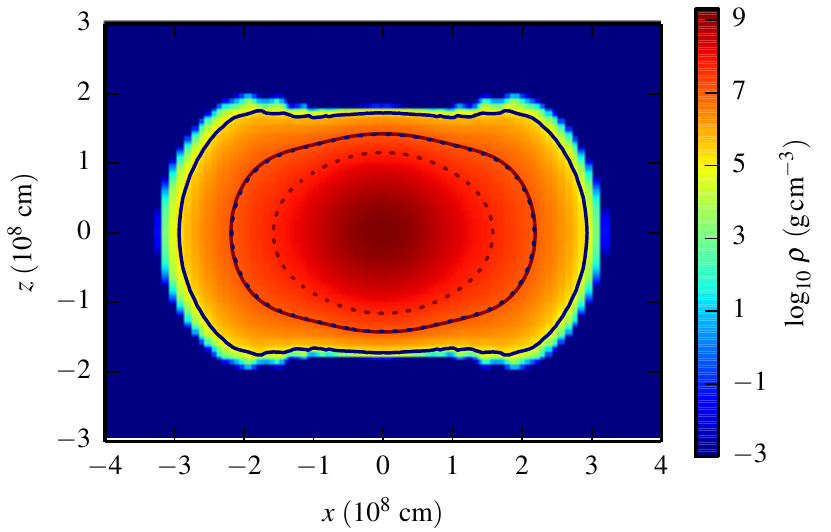}\\[0.7em]
  (c) AWD$3$, $M = 2.00~\msol$\\
  \includegraphics[width=6.8cm]{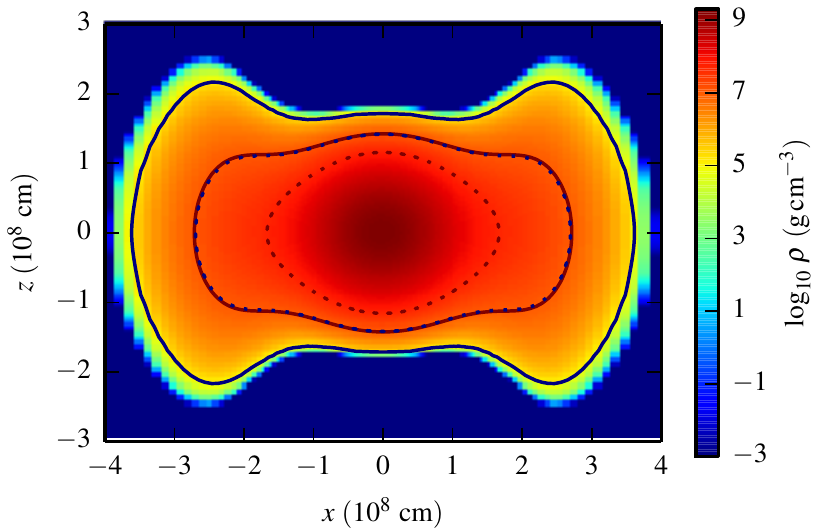}
  \caption{Initial rotation laws and density distributions of all our
    rapidly rotating progenitor models. The lower three panels display
    cross-sections of the WDs with the rotational axis in
    $z$-direction. The density is color coded.  Solid contours mark
    density levels of $\rho = 10^5$ and $10^7\,\gcc$, corresponding to
    the transition densities between the main burning stages
    (producing iron group and intermediate mass elements,
    respectively, or ceasing burning) in our detonation yield
    tables. For comparison, the dotted contours at
    $\rho = 1.047 \times 10^7$ and $5.248 \times 10^7\,\gcc$ show the
    transition densities assumed in the earlier study by
    \citet{pfannes2010b}.}
  \label{fig:msetup_inimod}
\end{figure}

The angular velocity distributions have central values
$\Omega_\text{c}$ between $1.6$ and $4.6\,\rads$ and peak values
$\Omega_\text{peak} \sim 5\,\rads$.  Shapes of the rotation laws and
the initial density profiles are illustrated in
Fig.~\ref{fig:msetup_inimod}.
The lower three panels show $x$--$z$-slices through the WDs along the
rotational axis, which is in the $z$-direction, and the upper panel
displays the angular velocity as a function of the radial coordinate
$r = (x^2 + y^2)^{1/2}$, where we have assumed constant angular
velocity on cylinders around the axis of rotation (``barotropic
rotation''). \citet{yoon2004a} attribute the general shape of the
angular velocity distribution with a maximum at around 1,000 to
1,500\,km to the fact that as the mass increases toward explosion the
more slowly rotating inner core contracts faster than the rapidly,
nearly critically rotating surface layers and, moreover, to the limit
they impose on the gain of angular momentum by accretion.

Since during the simmering phase prior to the explosion convection
sets in (see, e.g., \citealt{nonaka2012a}), the innermost few hundred
kilometers of the core may be in rigid rotation by the time the
deflagration phase starts. In order to account for this effect, we
assumed constant angular velocity for the inner 600~km in model
AWD4. We note, however, that the influence of rapid rotation on the
convective simmering phase is still unexplored.
  
At the highest densities, the WD cores do not deviate
much from spherical symmetry, whereas at lower densities the stars show an
extended ``bulge'' of material in equatorial direction that becomes
more pronounced with increasing total angular momentum and total mass.

For completeness, Table~\ref{tab:msetup_inimod_param} provides several
additional quantities: $r_\text{equator}$ and $r_\text{pole}$ are the
equatorial and polar radii and the ratio
$r_\text{equator} / r_\text{pole}$ is the eccentricity parameter.
$E_\text{grav}$, $E_\text{int}$, $E_\text{rot}$, and $E_\text{bind}$
are the initial values of gravitational potential, internal,
rotational kinetic, and effective binding energy
($E_\text{bind} = E_\text{grav} + E_\text{int} + E_\text{rot}$),
respectively.  Finally, $\beta = E_\text{rot} / |E_\text{grav}|$ is
the ratio of rotational energy and gravitational binding energy.

\subsection{Explosion scenarios}
\label{sec:msetup_expl_scen}

For all three AWD models, we have investigated the delayed-detonation
scenario, which is one of the currently favored scenarios for
modeling normal SNe~Ia (models AWD$i$ddt) and which is
the main subject of this article. In this
scenario, it is assumed that the supernova explosion starts as a
turbulent deflagration in the central core and later turns into a
detonation. In addition, for the AWD3 model, we also investigated a
pure deflagration (model AWD3def), which would be the outcome if the
delayed detonation fails to trigger, and a centrally initiated
spontaneous detonation (model AWD3det) to test the impact of the
changes in the detonation scheme used in our work in comparison to
\citet{pfannes2010b, pfannes2010a} (see also
Sect.~\ref{sec:nmethods_hydro} for more details). 

We begin with a discussion of the delayed-detonation models. As was
done previously for the deflagration phase of non-rotating WDs
\citep[e.g.][]{niemeyer1996a,reinecke2002d,roepke2006a,roepke2007c}
our rotating models are ignited in multiple ignition kernels around
the center. Although recent studies of the pre-ignition convection
phase in non-rotating Chandrasekhar-mass WDs do not support multispot
ignition \citep{nonaka2012a}, the situation for rotating progenitors
is less clear \citep{kuhlen2006a}. Our motivation for assuming a
pronounced multispot ignition scenario is to maximize the
deflagration strength and the pre-expansion when the DDT is
triggered. As noted already by \citet{steinmetz1992a}, pure
detonations produce almost pure iron-group elements (IGEs) in the
ejecta and virtually no intermediate-mass elements (IMEs), which is a
potential problem for reproducing observed features of SNe~Ia. This
will be checked with our radiative transfer simulations. In our
delayed detonation model, we aim at maximizing the IME production to
test the capabilities of the model. Therefore, we put $1600$ ignition
kernels in a spherical volume of radius $180$~km around the center at
exactly the same positions as in previous studies (the model of
\citealp{roepke2007c} and model N1600Cdef in \citealt{fink2014a}; see
Fig.~\ref{fig:sresults_evol}). However, different kernel radii of
$6$~km (in between those of the two mentioned studies) are chosen.

For the transition to a supersonic detonation, a suitable criterion
has to be chosen. In one-dimensional simulations, DDTs are
parametrized relatively arbitrarily.  A commonly used scheme is to
trigger a detonation when the deflagration flame reaches fuel
densities $\rho_\text{u} \lesssim 10^7\,\gcc$. However, this criterion
is only motivated by optimizing the agreement of synthetic with real
observables of SNe~Ia or, alternatively, by matching the
nucleosynthesis yields \citep{hoeflich1996a, iwamoto1999a}. A rigorous
determination of the DDT parameters from first principles is not
possible, as the exact physical mechanism leading to a DDT is still
unknown. An improvement in this sense is achieved by multidimensional
modeling, as turbulent flame propagation is an inherently
three-dimensional process.  This allows a physically motivated
parametrization of the DDT from properties of the flame (which itself
is only a parametrization in one-dimensional modeling).  There exist
different DDT criteria, that are related to the onset of the
distributed burning regime. The distributed burning regime is reached
when at low densities the flame becomes sufficiently thick so that
small eddies start to penetrate into the reaction zone and mix hot ash
with cold fuel without immediately burning \citep[cf.][]{peters2000a}.
This may lead to conditions that favor the formation of a detonation
via the Zel'dovich gradient mechanism \citep{zeldovich1970a} or
related processes \citep[e.g.][]{seitenzahl2009b, woosley2009a}.  An
equivalent condition for entering this regime is
\begin{equation}
  \label{eq:distr-burning}
  \mathit{Ka} \gtrsim 1,
\end{equation}
with
\begin{equation}
  \label{eq:karlovitz-number}
  \mathit{Ka} = \left(\frac{\delta_\text{lam}}{l_\text{Gibs}}\right)^{1/2}
\end{equation}
being the Karlovitz number.  Here, $\delta_\text{lam}$ is the
laminar flame width and $l_\text{Gibs}$ is the Gibson scale (the
size of the eddy that turns over in a laminar flame crossing time).
\citet{golombek2005a} and \citet{roepke2007b} used
\eqref{eq:distr-burning} as criterion for the DDT in two- and
three-dimensional simulations, following earlier suggestions by
\citet{niemeyer1997b}.  However, \citet{woosley2007a} argues that at
the onset of distributed burning the first structures to form are too
small to detonate and yet more mixing, lower densities, and higher
Karlovitz numbers are required.\footnote{Strictly speaking, within the
  distributed burning regime only an effective Karlovitz number can be
  defined because the concept of a laminar flame width does not exist
  there.}
Therefore, \citet{kasen2009a} chose critical
values of the Karlovitz number between $250$ and $750$, which causes a
somewhat later transition.  In this study, similar criteria are used.
A DDT is initiated when at the flame the criteria
\begin{equation}
  \label{eq:ddtcrit}
  \begin{split}
  &\mathit{Ka} \ge 250 \quad \text{and}\\
  &6 \times 10^6~\gcc < \rho_\text{u} < 1.2 \times 10^7~\gcc
  \end{split}
\end{equation}
are fulfilled. Consequently, transitions can (and will) happen in
different locations and at different times in the exploding WD\@.  In
our models, in the final detonation phase almost all the C/O of the WD
is burned to iron-group and intermediate-mass elements. However, in
detail the outcome is not very sensitive to the details of the
transition criterion we assume.\footnote{The DDT criterion
  employed here corresponds to that used by \cite{kasen2009a} whereas
  the more elaborate criterion of \citet{ciaraldi2013a} was used in
  the study of \citet{seitenzahl2013a}.}

In our prompt detonation model the detonation is initiated by setting
the level set function to positive values inside a small central
volume with a radius of $7$~km. The further evolution is computed by
means of the method described in detail in
Sect.~\ref{sec:nmethods_hydro} and in
Sect.~\ref{sec:discussion_det}. The detonation scheme ensures that the
detonation speed, the energy release, and the nucleosynthesis are
consistently computed. Finally, in the pure deflagration model AWD3def
we use identical initial conditions as in our delayed detonation model
AWD3ddt, but we assume that a deflagration-to-detonation transition
does not happen. Simulation data for all models are available via the
Heidelberg Supernova Model Archive (HESMA, \citealt{kromer2017a}).

\section{Numerical methods}
\label{sec:nmethods}

The numerical methods we apply here are very similar to those used by
\citet{pfannes2010b, pfannes2010a}. In this section we summarize these
methods and describe improvements with respect to the previous
studies \citep[see also][]{seitenzahl2013a}.

\subsection{Hydrodynamics}
\label{sec:nmethods_hydro}

Our simulations are carried out in three dimensions with the finite
volume hydrodynamics code \textsc{leafs} (\citealt{reinecke2002b};
most updates described in \citealt{seitenzahl2013a} are included). We
use a ``hybrid'' moving grid as developed by \citet{roepke2005c, roepke2006a} with
$512^3$ cells: while an inner uniform part of the grid (initial
spatial resolution: 1.9~km) tracks the deflagration flame, an outer
part with exponentially growing cell sizes tracks the overall
expansion of the ejecta. All explosion models are followed until $t =
100$\,s, when homologous expansion is reached to a good approximation.

As our initial WD models significantly depart from spherical symmetry,
the gravitational potential has to be calculated more accurately than
in previous studies. Here, we adopt the multipole gravity solver from
\citet{pfannes2010b, pfannes2010a} but, in contrast to
\citeauthor{pfannes2010a}, we do not assume rotational symmetry
($m = 0$) of the mass distribution, but include the full
quadrupole term ($l = 2$, $m \le l$).

Reaction fronts of explosive thermonuclear burning are approximated as
infinitesimally thin discontinuities between fuel and ash. These
discontinuities are tracked independently for deflagrations and
detonations using a level set technique (see \citealt{reinecke1999a},
and references therein; \citealt{golombek2005a}).  The composition of
the matter is approximated by five species: helium, carbon, oxygen and
representative species for both intermediate-mass and iron-group
elements. Changes in the composition and the release of nuclear
binding energy are assumed to occur instantaneously behind the burning
front.  The final composition of the ashes and with it the reaction
$q$-values are taken (as function of fuel density) from pre-calculated
tables (CO detonation: see Fig.~A.1 of \citealt{fink2010a}; CO
deflagration: see Fig.~A1\,b of \citealt{fink2014a}) that were
iteratively calibrated to our large post-processing nuclear reaction
network (see Sect.~\ref{sec:nmethods_pp}). 
This is one of the main improvements with respect to
\citet{pfannes2010b, pfannes2010a}, who used much coarser
approximations in this respect.  Everywhere on the grid, the
composition in nuclear statistical equilibrium (NSE) and the electron
fraction $Y_\text{e}$ are adjusted according to the thermodynamic
background state \citep[see][]{seitenzahl2009a}.

The burning speeds are also determined as functions of local quantities
on the grid as described in \citet{seitenzahl2013a}: an effective
deflagration speed on the grid-scale is determined on the basis of a
localized subgrid model \citep{schmidt2006b, schmidt2006c} which takes
into account turbulence on unresolved scales. The detonation speed is
taken from a table as described in \citet{fink2010a}: at high
densities (${\ge} 10^7\,\gcc$) the detonation is assumed to be
pathological \citep{sharpe1999a,gamezo1999a}; at low densities (${<} 10^7\,\gcc$) a
Chapman--Jouguet like velocity is determined for the incomplete
burning regime.

\subsection{Nucleosynthesis post-processing}
\label{sec:nmethods_pp}

We distribute $10^6$ equal-mass tracer particles in the asymmetric
initial WD models using a rejection method \citep[see
e.g.][]{press2007a} to properly sample the asymmetric mass
distribution. During the hydrodynamic simulations, the tracer
particles are passively advected with the flow and record a Lagrangian
representation of the explosion. Detailed nucleosynthetic yields are
then calculated by solving a large nuclear reaction network consisting
of 384 species (ranging up to $^{98}$Mo; see \citealt{travaglio2004a})
for all tracer trajectories. The reaction rates we used were taken
from an updated version of the RE\-AC\-LIB library \citep[][updated
2009]{rauscher2000a}.

\subsection{Radiative transfer}
\label{sec:nmethods_rt}

For our time-dependent 3D Monte Carlo radiative transfer calculations
with the \textsc{artis} code \citep{kromer2009a, sim2007b}, the final
ejecta density and the detailed post-processing abundances are mapped
on a $50^3$ Cartesian grid using the scheme described in
\citet{fink2014a} and \citet{kromer2010a}. In each radiative transfer
simulation we use $10^8$ photon packets and follow their evolution for
111 logarithmically spaced time steps between 2 and 120~d after the
explosion. We apply the atomic data set as described by
\citet{gall2012a}, use a gray approximation in optically thick
cells \citep[cf.][]{kromer2009a}, and assume local thermodynamic
equilibrium at early times, that is, $t < 3$~d.

\section{Simulation results}
\label{sec:sresults}

\subsection{The prompt detonation model AWD3det}
\label{sec:sresults_det}

As was mentioned before, this model was computed to test the impact
of the changes we made in our computational scheme as compared to
\citet{pfannes2010b, pfannes2010a}.

\begin{figure}
  \centering
  \includegraphics[width=\columnwidth]{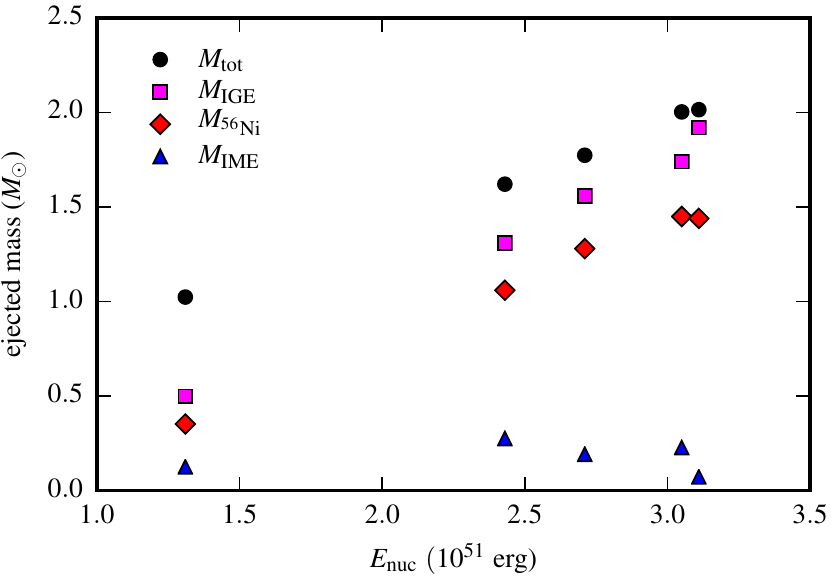}
  \caption{Integrated nucleosynthetic yields of all explosion models.
   The models are (from left to right): AWD3def, AWD1ddt, AWD4ddt,
   AWD3ddt, and AWD3det. Black dots depict the total ejecta masses.
   The colored symbols are (from top to bottom): iron-group elements
   (IGE), \nni\, and intermediate-mass elements (IME).}  
   \label{fig:sresults_integr_yields}
\end{figure}

After the central ignition, the detonation burns supersonically
through the whole WD without giving it time to expand. Thus, the
nucleosynthesis products of the thermonuclear burning
depend only on the initial density as extracted from 
our detonation-yield table (see Fig.~A.1 of \citealt{fink2010a}):
IGEs are produced for $\rho \gtrsim 10^7\,\gcc$ and IMEs for $10^7\,\gcc
\gtrsim \rho \gtrsim 10^5\,\gcc$.  Due to the initial density
distribution of the AWD3 model (the solid contour lines in
Fig.~\ref{fig:msetup_inimod} mark the transition densities), this
leads to a very high mass of IGEs (1.92\,\msol, 1.44\,\msol\ of it \nni;
see Table~\ref{tab:sresults}
\begin{table*}
  \centering
  \caption{Results}
  \label{tab:sresults}
  \begin{tabular}{cccccccc}
    \hline
    \hline
    & & AWD1ddt & AWD4ddt & AWD3ddt & AWD3det & \multicolumn{2}{c}{AWD3def} \\
    \hline
    & & & & & & ejecta & remnant object \\
    $t_\text{DDT}$ & (s) & $0.857$ & $0.806$ & $0.818$ & -- & -- \\
    $\Delta t_\text{DDT}$ & (s) & $0.146$ & $0.144$ & $0.125$ & -- & -- \\
    $N_\text{DDT}$ & & $215$ & $152$ & $133$ & -- & -- \\
    $E_\text{nuc}^\text{DDT}$ & $(10^{51}~\text{erg})$ & $0.563$ & $0.468$ & $0.484$ & -- & -- \\
    $E_\text{nuc}$ & $(10^{51}~\text{erg})$ & $2.43$ & $2.71$ & $3.05$ & $3.11$ & $1.31$ \\
    $E_\text{nuc}/|E_\text{bind}|$ & & $3.4$ & $3.0$ & $2.8$ & $2.9$ & $1.2$ \\
    $E_\text{tot}$ & $(10^{51}~\text{erg})$ & $1.71$ & $1.83$ & $1.97$ & $2.03$ & $0.246$ \\
    $M_\text{tot}$ & $(\msol)$ & $1.62$ & $1.77$ & $2.00$ & $2.02$ & $1.02$ & $0.980$ \\
    $M_\text{IGE}$ & $(\msol)$ & $1.31$ & $1.56$ & $1.74$ & $1.92$ & $0.499$ & $0.117$ \\
    $M_{\nni}$ & $(\msol)$ & $1.06$ & $1.28$ & $1.45$ & $1.44$ & $0.353$ & $8.40\times10^{-2}$ \\
    $M_\text{IME}$ & $(\msol)$ & $0.276$ & $0.193$ & $0.228$ & $7.32\times10^{-2}$ & $0.126$ & $8.00\times10^{-2}$ \\
    $M_{\nox}$ & $(\msol)$ & $3.07\times10^{-2}$ & $2.18\times10^{-2}$ & $2.84\times10^{-2}$ & $1.01\times10^{-2}$ & $0.220$ & $0.407$ \\
    $M_{\ncarb}$ & $(\msol)$ & $6.16\times10^{-4}$ & $4.28\times10^{-4}$ & $8.16\times10^{-4}$ & $4.53\times10^{-4}$ & $0.162$ & $0.342$ \\
    \hline
  \end{tabular}
\end{table*}
and Fig.~\ref{fig:sresults_integr_yields})
and only a low mass of 
IMEs (0.07\,\msol). We also note the rather high
polar velocity of radioactive nickel and IMEs in this model (fourth row
in Fig.~\ref{fig:sresults_vdistr_yields}). In general terms, our
results are in fair agreement with \citet{pfannes2010b} although there
are some differences which, however, can be understood (see Sect.~\ref{sec:discussion}).

\subsection{Delayed detonation models}
\label{sec:sresults_deldet}

\begin{figure}
  \centering
  \includegraphics[width=5.75cm]{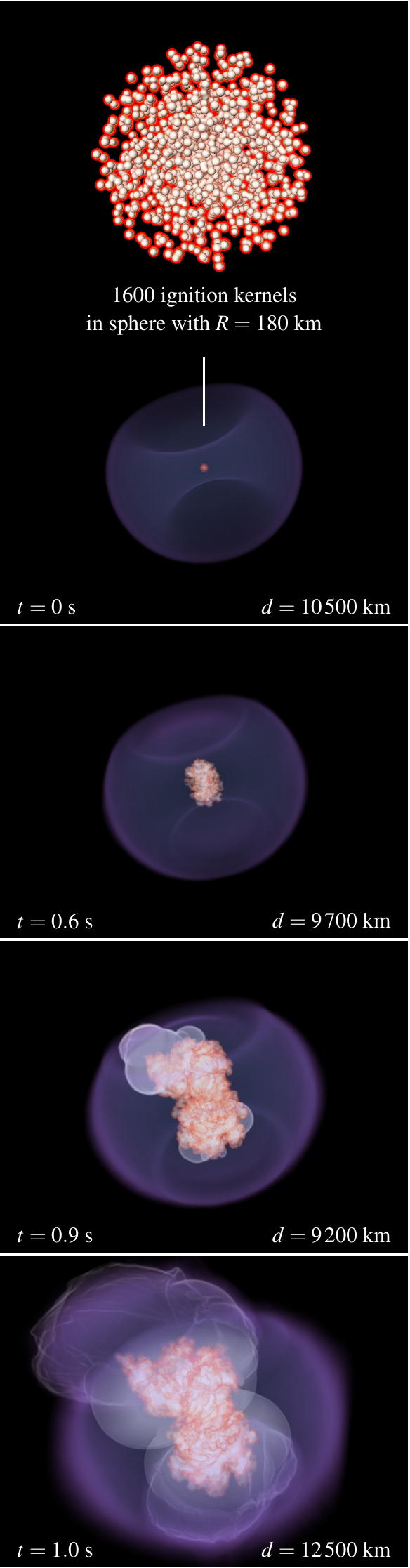}
  \caption{Temporal evolution of the explosion for the AWD3ddt
    model. The top panel shows the ignition configuration ($t=0$\,s)
    with a zoom-in on the ignition kernels. The lower panels show the
    evolution of the deflagration ash (pink) and the detonation front
    (light blue). The dark blue surface indicates the outer boundary
    of the star. $d$ denotes the size of the
    computational domain. The decrease in $d$ is due to
    our nested-grid approach \citep{roepke2006a}.}
  \label{fig:sresults_evol}
\end{figure}

Despite the roughly spherically symmetric ignition in multiple spots,
the initial deflagration flame develops a pronounced anisotropy in all
models.  The flame propagation is much faster along the rotational axis (the
$z$-axis) than perpendicular to it (see Fig.~\ref{fig:sresults_evol}
which shows the evolution of the explosion for model AWD3ddt). This 
behavior has been shown
to be characteristic for deflagrations in rapidly rotating WDs
(\citealt{pfannes2010a}).
According to \citeauthor{pfannes2010a} this has two reasons: first,
the effective gravitational acceleration and with it the buoyancy
force is larger along the rotational axis than perpendicular to it
(due to a steeper density decline and a minimum of the centrifugal
acceleration along the rotational axis); secondly, and more important,
the matter flow perpendicular to the rotational axis and thus 
turbulence-induced flame acceleration is suppressed due to an
effective angular momentum barrier.

Owing to its asymmetric propagation, the deflagration flame reaches
the low-density edge of the star close to the poles while the matter
in equatorial directions is still mostly unburned. Our DDT criterion
Eq.~\eqref{eq:ddtcrit} is first met at the trailing edges of the outer
deflagration flame front, where the flame brush is wider due to the
lower densities. Consequently, this is where the distributed-burning
regime is reached first \citep[cf.][]{roepke2007b, roepke2007d}.

For all our models, the DDT occurs at times $t_\text{DDT} \sim
0.80$--$0.85$\,s (see Table~\ref{tab:sresults}). Our approach allows
for multiple transition points and, indeed, we find around
$100$--$200$ such spots where Eq.~\eqref{eq:ddtcrit} is met within a
time interval $\Delta t_\text{DDT} \sim 0.15$\,s. Detonations are
initiated at similar times both close to the north and the south pole
of the WD\@.  From there, they propagate toward the equatorial plane
and rapidly consume the remaining unburned fuel. Collisions of the
leading shocks cause some enhanced compression when both detonation
fronts meet close to the equatorial plane. This can still be seen in
the final distribution of \nsi\ and \nox\ in
Fig.~\ref{fig:sresults_vdistr_yields}.  Moreover, since during the
deflagration phase only moderate pre-expansion happened, the
detonation can still produce significant amounts of IGEs. This differs
from model N1600C of \citet{seitenzahl2013a}, which uses the same
ignition setup as our models here. However, with a non-rotating
progenitor star the pre-expansion during the deflagration is much more
efficient and the ensuing detonation produces mainly IMEs and O.

Our initial models form a series of increasing rotational kinetic energy
$E_\text{rot}$ and total mass $M$ of the progenitor (in the order
AWD1, AWD4, AWD3). More massive models are more tightly bound and
therefore start at lower initial total energy ($E_\text{tot} =
E_\text{grav} + E_\text{int} + E_\text{kin}$; see
Table~\ref{tab:sresults} for explosion energetics results), despite
the larger $E_\text{rot}$ (see Table~\ref{tab:msetup_inimod_param}).
However, due to the increasing amount of mass being burned at high
densities (corresponding to an increasing nuclear energy release
$E_\text{nuc}$) within the series, all models become unbound rapidly and
the more massive models even reach larger final $E_\text{tot}$ and
$M_\text{IGE}$ (see also Fig.~\ref{fig:sresults_integr_yields}):
$M_\text{IGE}$ ranges from $1.31$ to $1.74\,\msol$.  Due to the rapid
unbinding and expansion of the star, $E_\text{kin}^\text{asymp.}
\approx E_\text{tot}(t = 100\,\text{s})$ holds for all delayed
detonation (and also the pure detonation) models.
The corresponding \nni\ masses are in the range $1.06$--$1.45\,\msol$.

Compared to the pure detonation model AWD3det, the delayed-detonation
model AWD3ddt has lower IGE mass due to the expansion of the star
during the deflagration phase. Despite the chosen vigorous ignition
setup, the differences are, however, relatively limited. This is in
contrast to non-rotating models, in which the strength of the
deflagration is distinctly anticorrelated with the mass of IGEs
\citep{roepke2007b}: a weak deflagration implies that more unburned
material is available for the detonation and, thus, more IGEs (and
\nni) are produced. In the fast rotating models the deflagration wave
propagates predominantly along the axis of rotation, which results
always in lots of unburned fuel at the onset of the
detonation. Remarkably, the AWD3ddt model even slightly surpasses
AWD3det in \nni\ mass due to its lower degree of
neutronization. Compared to the (rotating) pure detonation model, the
delayed-detonation models produce more IMEs. However, the total IME
masses are still relatively small ($0.2$--$0.3\,\msol$) and the exact
amounts depend on the details of the initial density distributions. In
all models, the amounts of unburned \ncarb\ and \nox\ are very small and
restricted to the outermost layers.

\subsection{The pure deflagration model AWD3def}
\label{sec:sresults_defl}

In order to determine the observable outcome of the explosion of a
rapidly rotating WD in the case that the DDT fails, we have calculated a
pure deflagration version with AWD3 as the initial model.  As in
\citet{pfannes2010a} the deflagration evolves preferentially along the
rotation axis of the progenitor WD and leaves large fractions
(${\sim}50\%$ of the mass) of the star unburned (c.f.,
Fig.~\ref{fig:sresults_vdistr_yields}, bottom row). Compared to the
other models of our sample, the energy release is relatively low and
just surpasses the initial binding energy of the WD: $E_\text{nuc} =
1.31 \times 10^{51}\,\text{erg} \approx 1.2 \, |E_\text{bind}|$. As
discussed in detail for some of the models in \citet{fink2014a}, also
here significant amounts of the WD's matter do not reach escape
velocity and form a gravitationally-bound remnant object of
$0.98\,\msol$ after the explosion.\footnote{In model N1600Cdef of
  \citet{fink2014a} the progenitor star is less tightly
  bound. Therefore, the deflagration fully unbinds the progenitor WD
  and $E_\text{nuc}/|E_\text{bind}| \approx 2.1$. In our model
  AWD3def, the deflagration fails to unbind the WD although we use the
  same vigorous ignition configuration as in N1600Cdef.}  A cell on
our numerical grid is considered as part of the remnant, if it has a
negative value of the specific asymptotic kinetic energy,
$\epsilon_\text{kin,a} = \epsilon_\text{grav} + \epsilon_\text{kin}$,
at the end of the simulation. Here, $\epsilon_\text{grav}$ and
$\epsilon_\text{kin}$ are the specific gravitational and kinetic
energies.  The internal energy was found to be negligible for this
calculation. As a result of the incomplete disruption of the star, a
mere $0.353\,\msol$ of \nni\ and $0.126\,\msol$ of IMEs are ejected in
the explosion. Remarkably, $12\%$ of the synthesized IGEs and $9\%$ of
IMEs become part of the bound remnant. We do not find that the remnant
object receives any significant kick from the explosion ejecta.

\subsection{The distribution of the nucleosynthesis products in
  velocity space}
\label{sec:sresults_vdistr_yields}

An overview of the integrated nucleosynthetic yields of our models is
given in Fig.~\ref{fig:sresults_integr_yields} and in
Table~\ref{tab:sresults}.  In this section, we describe the details of
the yield distribution in asymptotic velocity space, which is also
crucial for the predictions of observables from the models described in
Sect.~\ref{sec:synth_obs}.

\begin{figure*}
  \centering
  \includegraphics[width=17cm]{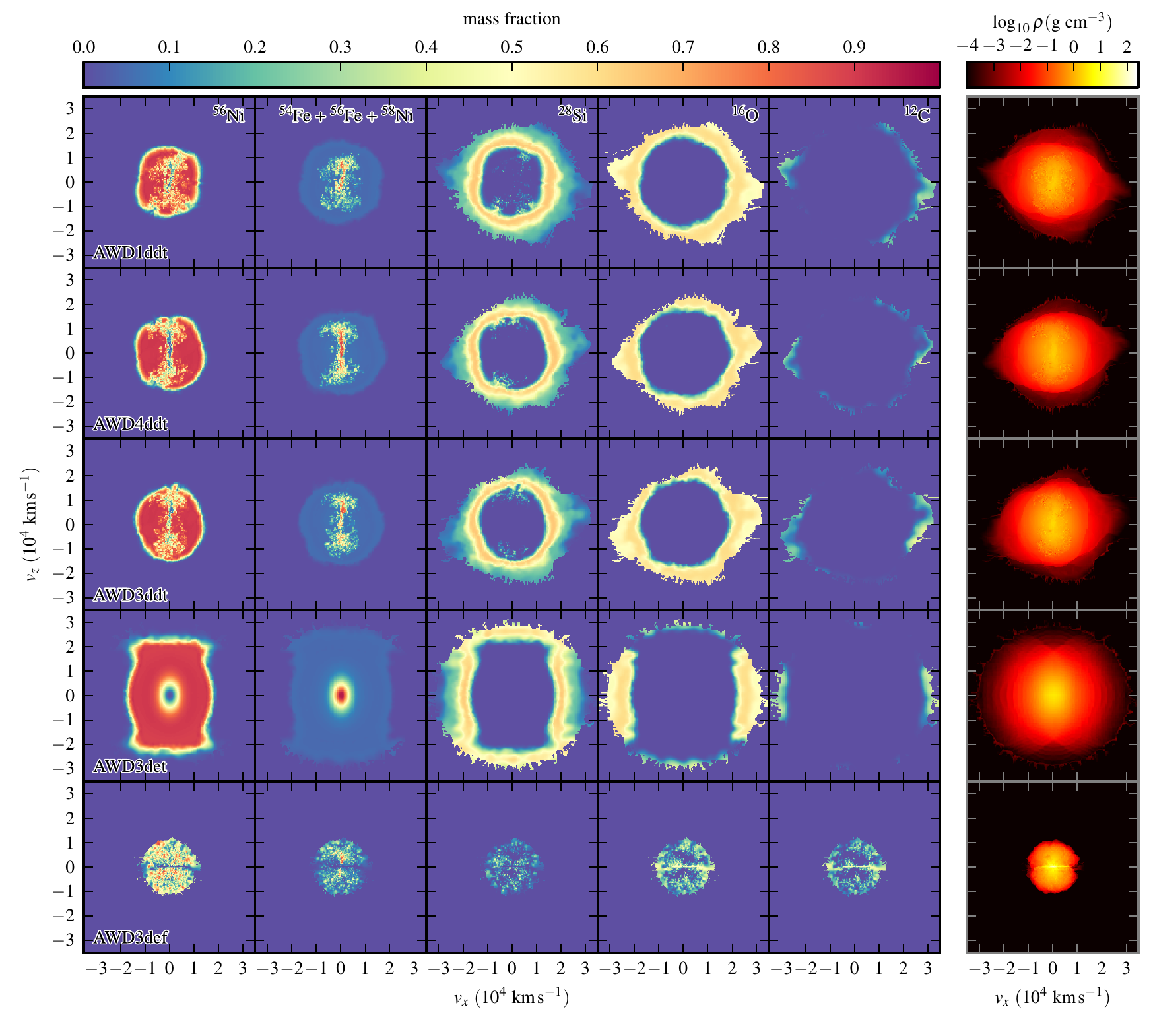}
  \caption{Abundance distributions and densities of the ejecta of all
    models (top to bottom) at $t = 100$\,s.  Shown are slices through
    the $x$--$z$-plane in velocity space for $^{56}\text{Ni}$,
    $^{54}\text{Fe} + {^{56}\text{Fe}} + {^{58}\text{Ni}}$,
    $^{28}\text{Si}$, $^{16}\text{O}$, $^{12}\text{C}$ and $\log_{10}
    \rho$ (from left to right).}
  \label{fig:sresults_vdistr_yields}
\end{figure*}

In Fig.~\ref{fig:sresults_vdistr_yields}
we show in velocity space two-dimensional slices through the final
abundance and density structures of the ejecta of all models (for model
AWD3def, asymptotic velocities are used and thus the bound remnant is
excluded).  The abundance distributions were determined by mapping the
final post-processing results (see Sect.~\ref{sec:nmethods_pp}) on a $200^3$
Cartesian grid.

Model AWD3det shows a layered abundance pattern typical for pure CO
detonations.  The box-shaped distribution of \nni\ and other IGEs stem
from the break out of the detonation shock waves in polar direction,
which occurs significantly before all material in the equatorial plane
is burned \citep[cf.][]{steinmetz1992a}.

The complex abundance patterns of model AWD3def are very similar to
the pure deflagrations published by \citet{fink2014a} for non-rotating
progenitor WDs.  The asymmetric bipolar structure of the deflagration
(visible also in the deflagration phase of the corresponding DDT model
AWD3ddt in Fig.~\ref{fig:sresults_evol}) is evolving to a much more
symmetric structure in the free streaming phase, as the hot ashes near
both poles expand significantly in all directions, while the unburned
material close to the equatorial plane expands much less (this is also
reflected in the high density regions close to the equatorial plane).

The delayed detonation models show complex abundance patterns from the
deflagration phase in the inner regions that are surrounded by layered
structures of the detonation phase in the outer parts.  The
deflagration ashes have a different shape compared to the pure
deflagration model: they extend to higher velocities in both the $z$-
and the $-z$-direction and they do not extend as far into the regions
close to the equatorial plane.  These structures are caused by the
detonation shock waves emerging from both poles, which collide near
the equatorial plane at $t \sim 1.0$\,s.  The wave emerging from the
north pole then propagates through the ashes of the large southern
plume of deflagration ashes and vice versa.  In this way these ashes are
further accelerated in the polar direction and reach higher maximum
velocities than in pure deflagration models.  The former detonation
shocks are still visible in the final density structures at the upper
and lower edges and are the reason for the oblate shape of the outer
ejecta.  Asymmetric detonation initiation (away from the rotational
axis) in the DDT models is the reason why these oblate structures can
be tilted with respect to the equatorial plane.  

Models that involve a detonation phase have significantly higher
maximum ejecta velocities (${\sim}30\,000\,\kms$) than the pure
deflagration model (${\sim}10\,000\,\kms$).  This is a direct
consequence of the huge explosion energies of these models, which is
large compared to the initial binding energy, and consistent with
results obtained for non-rotating models (see \citealt{seitenzahl2013a,
  seitenzahl2014a}).

The central high density parts of the ejecta show a prolate asymmetry
in all models (also AWD3det and AWD3def).  This is caused by stronger
expansion toward the poles in the early explosion phases before
homologous expansion is reached.  For models that start with a
deflagration, the asymmetric flame propagation that also causes an
asymmetric expansion has already been discussed in
Sect.~\ref{sec:sresults_deldet}.  Another reason for stronger
expansion toward the poles that applies to all models is the
significantly earlier breakout of the burned matter in this direction
due to the much lower polar radius of all of our progenitor WD models.
This asymmetric expansion also causes a slightly asymmetric
distribution of the nucleosynthetic yields in final velocity space.
The velocities of the IMEs and the outer edges of the IGE-rich inner
parts tend to be slightly higher along the direction of the rotational
axis.  The impact of these asymmetries on the observable outcomes is,
however, only moderate (see Sect.~\ref{sec:synth_obs}).

\section{Synthetic observables}
\label{sec:synth_obs}

\begin{figure*}
  \sidecaption
  \includegraphics[width=12cm]{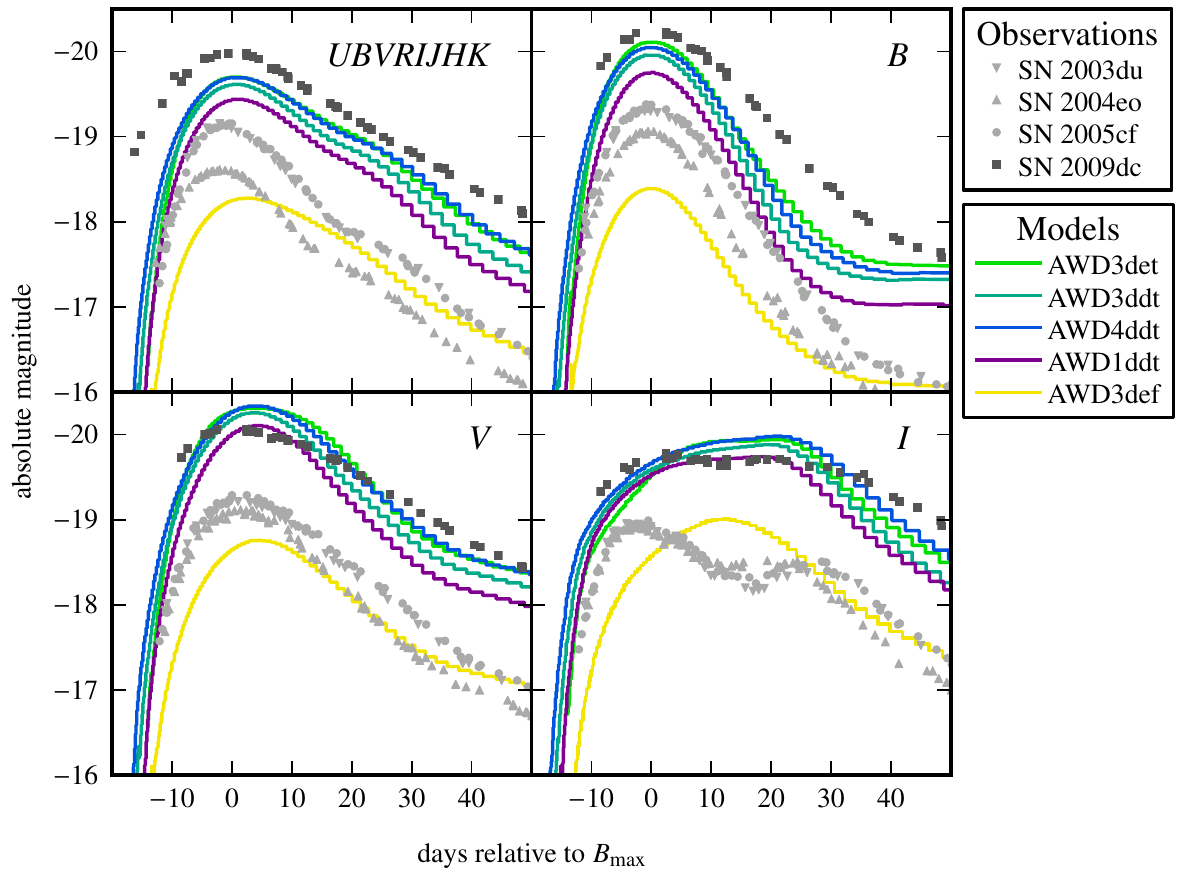}
  \caption{Angle-averaged synthetic lightcurves of our models,
    compared with the superluminous SN~Ia SN~2009dc (black squares)
    and ``normal'' SNe~Ia SN~2005cf (gray dots), SN~2004eo
    (gray triangles), and SN~2003du (gray inverted triangles).}
  \label{fig:synth_obs_lcs}
\end{figure*}

\begin{figure}
  \centering
  \includegraphics[width=\columnwidth]{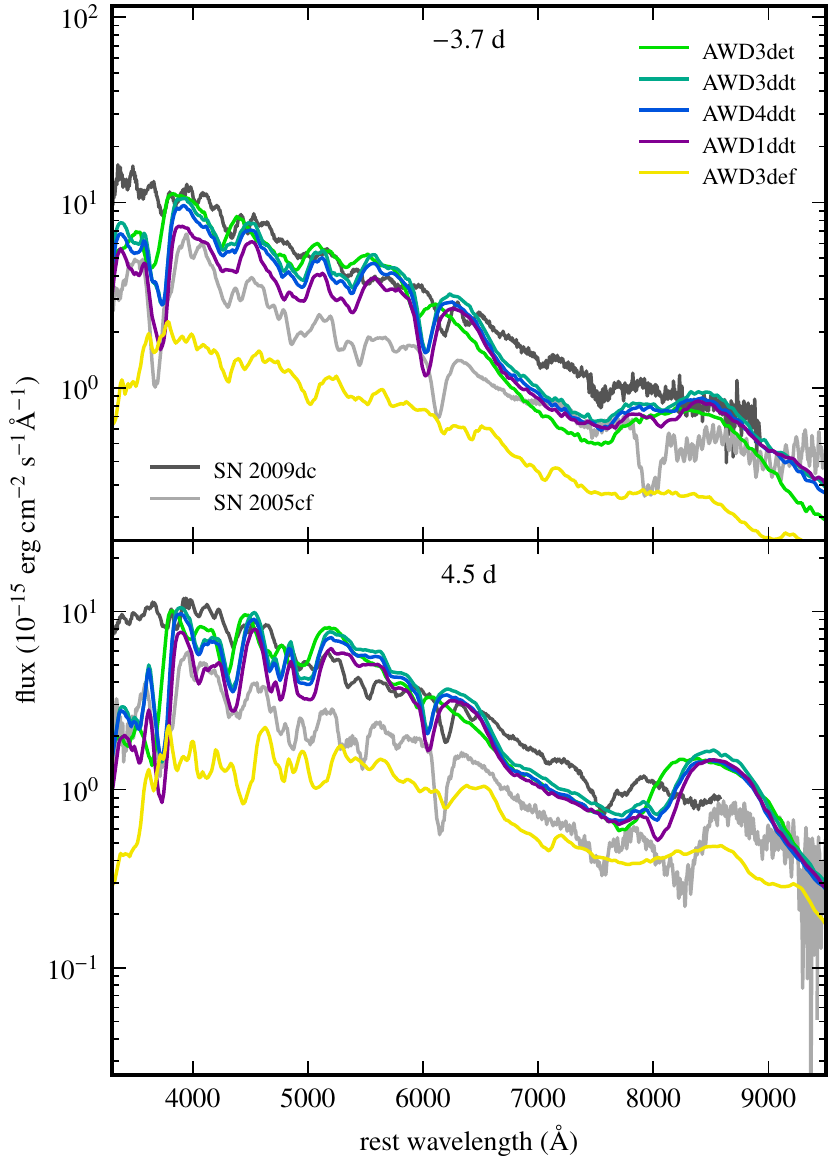}\\
  \caption{Angle-averaged synthetic spectra shortly before (upper
    panel) and after B-band maximum (lower panel). The spectra of a
    normal (SN~2005cf) and a superluminous (SN~2009dc) SN~Ia are
    shown in gray and black, respectively.}
  \label{fig:synth_obs_spectra}
\end{figure}

We have calculated synthetic observables for all explosion models
presented in Sect.~\ref{sec:sresults} with our Monte Carlo radiative
transfer code \textsc{artis} as described in
Sect.~\ref{sec:nmethods_rt}. The resulting lightcurves and spectra are
shown in Figs.~\ref{fig:synth_obs_lcs}
and \ref{fig:synth_obs_spectra}.
For comparison, we also show
observational data of normal and superluminous SNe~Ia.

From the bolometric lightcurves shown in the top-left panel of
Fig.~\ref{fig:synth_obs_lcs}, we conclude that our delayed and prompt
detonation models are brighter than normal SNe~Ia, but (at all times
shown) they are still not bright enough to explain the most luminous
observed SNe~Ia such as SN~2009dc. This is not too surprising since
none of our models reaches a \nni\ mass of $1.6\,\msol$, as suggested
for SN~2009dc by, for example, abundance-tomography studies of
\citet{hachinger2012a}. We also find that the bolometric lightcurves
of these models evolve significantly faster than observed lightcurves
of superluminous SNe~Ia, in particular during the rise.

A similar behavior is observed for band-limited lightcurves. For
example, we show $B$-, $V$- and $I$-band lightcurves in the other
panels of Fig.~\ref{fig:synth_obs_lcs}. In particular the $B$-band
lightcurves show a too fast past-maximum decline, but also the $V$-band
evolution of the models is significantly faster than observed in
superluminous SNe~Ia. This is a result of the high ejecta velocities
which, in turn, are inconsistent with observations.

Also in their synthetic spectra around maximum brightness
(Fig.~\ref{fig:synth_obs_spectra}) our delayed and pure detonation
models disagree with observations of superluminous SNe~Ia.  Most
importantly, the Si~\textsc{ii} $6355\,\AA$ feature shows a
significantly too large blueshift compared to observations.
This is the result of the large Si velocities of all of our models
discussed in Sect.~\ref{sec:sresults_vdistr_yields}.  A second failure
of all delayed detonation (as well as the prompt detonation) models
are the missing C~\textsc{ii} $6580\,\AA$ and $7234\,\AA$ lines which
are very prominent and persist up to two weeks past maximum in
superluminous SNe~Ia (\citealt{silverman2011a, taubenberger2011a,
  chakradhari2014a, parrent2016a}) but are not seen in our synthetic
spectra. This was to be expected since the models have very little
unburned material in their ejecta. Finally, the model spectra show a
significant lack of flux at wavelengths shorter than 4000\,\AA\
compared to superluminous SNe~Ia as, for example, SN~2009dc. This is
also reflected in the $B$-band lightcurves in
Fig.~\ref{fig:synth_obs_lcs}, where the models are systematically
fainter than SN~2009dc at all epochs. This indicates too much line
blanketing by IGEs in the model ejecta.

Owing to its significantly lower \nni\ mass, the pure deflagration
model AWD3def is not a candidate for superluminous SNe~Ia.  With a
bolometric peak magnitude of $\sim-18.3$ it is more than a magnitude
dimmer than the delayed and pure detonation models (see
Fig.~\ref{fig:synth_obs_lcs}). In fact, its bolometric peak is even
fainter than that of normal SNe~Ia. This is also reflected in the
band-limited lightcurves, where in particular $V$-band and bluer
lightcurves show a flux deficit compared to normal SNe~Ia. 
The peak magnitudes in the redder bands approach those of normal
SNe~Ia, but the singly-peaked lightcurves of the AWD3def model do not
match the observed double-peak structure in the NIR lightcurves of
normal SNe~Ia (see for example the $I$-band lightcurve in
Fig.~\ref{fig:synth_obs_lcs}).  
Synthetic spectra of model AWD3def around maximum brightness are shown
in Fig.~\ref{fig:synth_obs_spectra}. They show similar features as the
spectra of previously published pure deflagration models for
non-rotating progenitor WDs \citep[see e.g.][]{fink2014a,
  kromer2015a}.
In particular at early epochs the model spectra lack the strong
features of IMEs like Si, S or Ca, which are characteristic for normal
SNe~Ia. Instead, the spectra resemble those of Type~Iax SNe as found
for deflagrations in non-rotating WDs \citep{kromer2013a}.

\begin{figure*}
  \sidecaption
  \includegraphics[width=12cm]{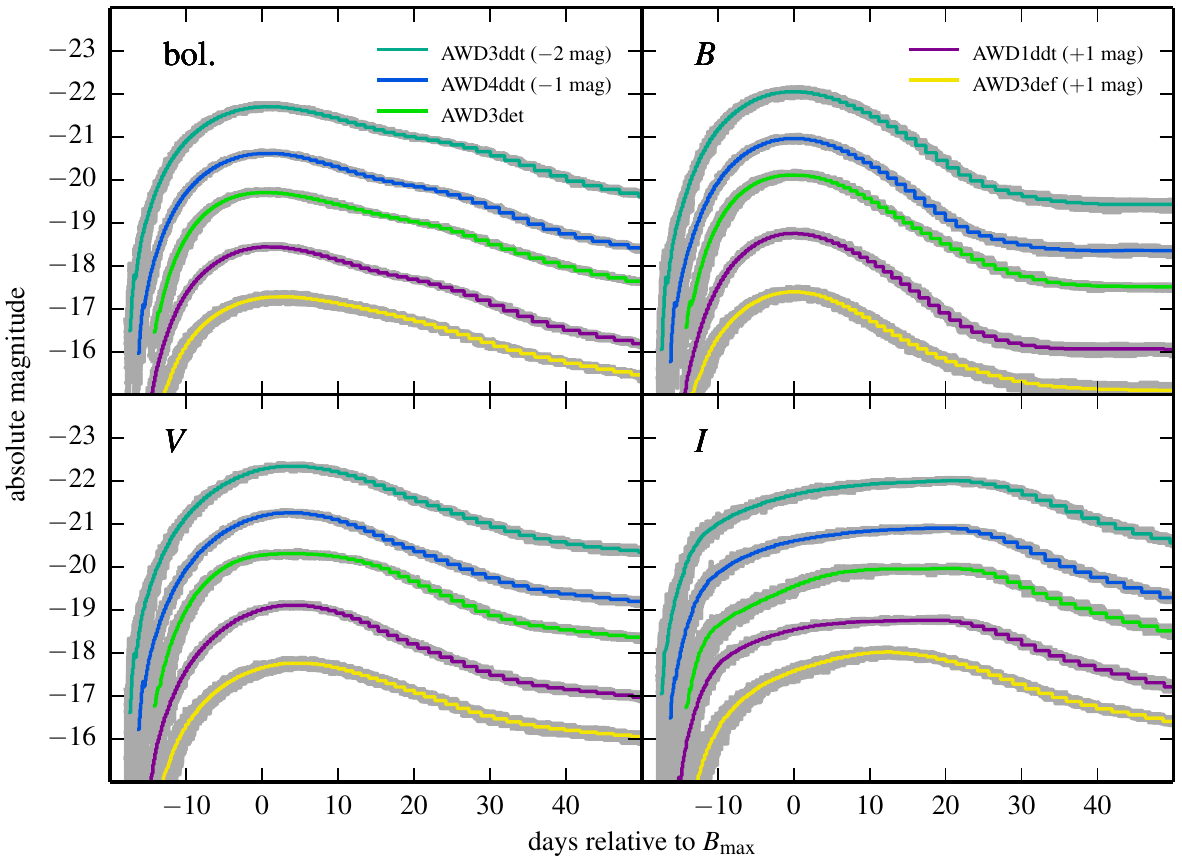}
  \caption{Bolometric and broadband lightcurves of all models for
    100 different viewing angles. The magnitudes are shifted as
    indicated. The time is given relative to $B$-band maximum.  As can
    be seen, the line-of-sight effects are small for all
    models.}
  \label{fig:los_lightcurves}
\end{figure*}

Despite some apparent ejecta asymmetries seen in
Fig.~\ref{fig:sresults_vdistr_yields} the synthetic lightcurves of all
our models show generally a very moderate viewing-angle dependence
(see Fig.~\ref{fig:los_lightcurves}),
which has the tendency to
decrease with time. This reflects the fact that the density structures
of the models are on large-scales fairly symmetric. The strongest
deviations from spherical symmetry in the density structure are
present in the outer layers, which are most important at early
times. In the innermost regions the ejecta show prominent asymmetries
in their chemical composition. These deeper layers are not yet visible
at the epochs studied here, since the ejecta are still largely
optically thick. However, at later epochs, when the ejecta become
optically thin, the compositional asymmetries could give rise to
characteristic emission features in nebular spectra.

\section{Discussion}
\label{sec:discussion}

\subsection{A self-consistent detonation scheme}
\label{sec:discussion_det}

As mentioned already in Sect.~\ref{sec:nmethods_hydro}, the main
improvement of the present numerical scheme for modeling detonations
over that used by \citet{pfannes2010b} is an energy release consistent
with the large post-processing nuclear network. The new scheme ensures
that the detonation speed, the energy release, and the nucleosynthesis
are consistent with each other. As a consequence,
\citeauthor{pfannes2010b} assumed significantly higher transition
densities between the main burning stages than those that result from
our self-consistent treatment (compare, e.g., the dotted and solid contours in
Fig.~\ref{fig:msetup_inimod}). For the prompt detonation of the AWD3
rotator this results in a significantly lower total energy release
$E_\text{nuc}$ of $2.65 \times 10^{51}\,\text{erg}$ in
\citet{pfannes2010b} as compared to our value of
$3.11 \times 10^{51}\,\text{erg}$, and a significantly lower IGE mass
of 1.75\,\msol\ from the post-processing as compared to our value of
1.92\,\msol.

\begin{figure}
  \centering
  \includegraphics[width=\columnwidth]{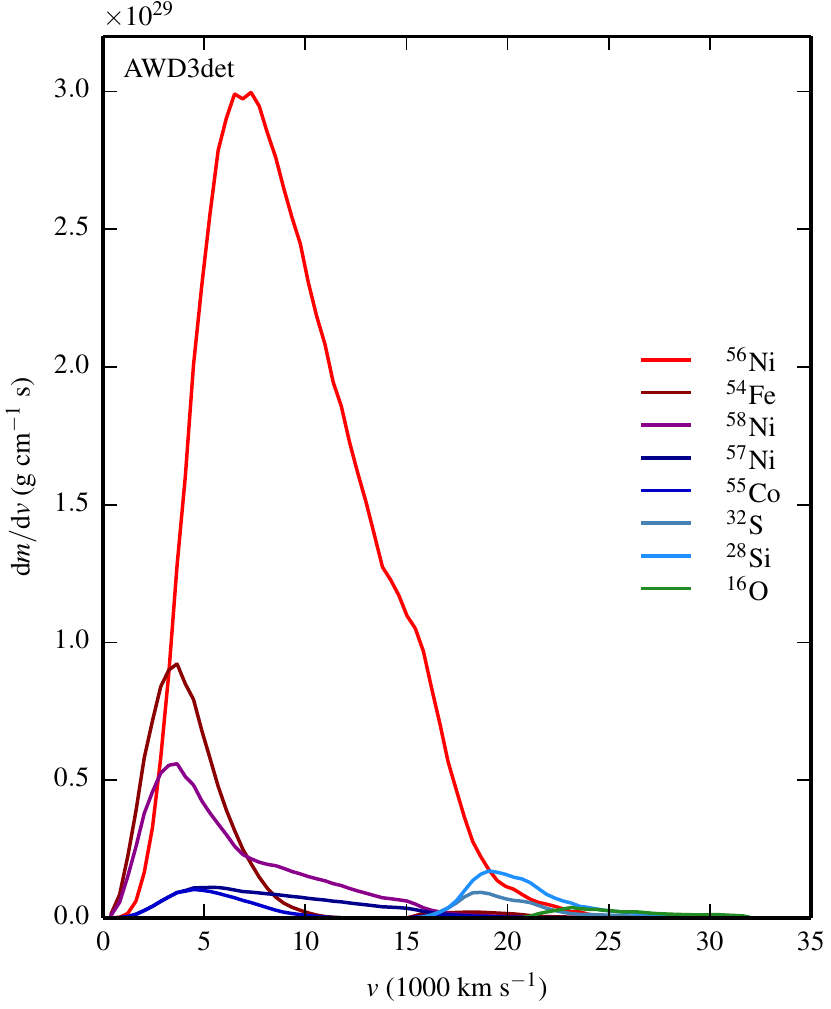}
  \caption{Angle-averaged velocity distribution of select nuclei in
    the ejecta of model AWD3det. The velocity bin size d$v$ is
    400\,$\kms$.}
  \label{fig:discussion_det_dmdv}
\end{figure}

Another major difference between our work and that of
\citet{pfannes2010b} concerns the velocity distribution of the
chemical elements synthesized in the explosion. Firstly, the higher
NSE transition density there naturally leads to a decrease of the
velocity at the interface between IGEs and IMEs. Secondly,
\citeauthor{pfannes2010b} calculate the final abundance profile in
velocity space with their reduced set of species, that is, with those
used in their hydrodynamic simulations and not those from the
post-processing step (corresponding to $M_\text{IGE} = 1.41$\,\msol\ --
referred to as ``high burning threshold'' or HBT in their paper).
This results in a further shift of IGEs to IMEs toward even lower
velocities.  Thus, in their Fig.~5b the abundances are inconsistent
with their detonation speeds and the energy release. In contrast to
their simulations, our model AWD3det has a maximum in the
angle-averaged silicon mass distribution at around $20\,000\,\kms$
(see Fig.~\ref{fig:discussion_det_dmdv}),
whereas \citet{pfannes2010b} find much lower values of about
$10\,000\,\kms$ only. 

We argue that our results are more realistic. \citet{pfannes2010b}
state that shear acting on the detonation front combined with the
cellular instability may shift the NSE burning threshold toward
higher densities and use the NSE transition densities for
deflagrations ($\rho_\text{NSE} \sim 5 \times 10^7\,\gcc$) as plausible
upper cutoff.  But, there are also effects acting in the opposite
direction (i.e.\ shifting $\rho_\text{NSE}$ to lower densities) such
as shock steepening in the density gradients of the WD star.  Thus, in
this work we use the detonation prescription from \citet{fink2010a}
instead, which resembles the planar detonation case.
With our improved modeling of burning in the detonation mode, we find
brighter pure detonation explosions than \citet{pfannes2010b}.

\subsection{Predicted observables and comparison with data}
\label{sec:discussion_synth_obs}

As far as their lightcurves and spectra are concerned, our rapidly
rotating delayed-detonation models are not too different from the
prompt detonations. They produce similar nickel masses, and also the
silicon expansion velocities, as seen in the blueshifts of the Si line
features in the spectra, are rather similar.  This can be understood
as follows. Since in our rapidly rotating delayed-detonation models
the asymmetric deflagration leaves behind major parts of the WDs
unburned before the DDT occurs, the explosions are always dominated by
the (short) detonation phase. Therefore, compared to prompt detonation
models, only slightly more IMEs are produced (due to the pre-expansion
phase before the detonation sets in). Consequently, the velocity of
the ejected IME material is very high as in the pure detonation
case. Thus, we find both prompt- and delayed-detonation models of
rapidly differentially-rotating super-\mch\ WDs to be incompatible
with observed properties of superluminous SNe~Ia. The models are not
bright enough, their lightcurves decline too fast in all wavebands,
and our synthetic spectra show blueshifts of the Si \textsc{ii}
$6355\,\AA$ absorption line feature that are far too high. High enough
\nni\ masses and low velocities of the Si-rich ejecta cannot be
realized at the same time in this kind of models. In addition, our
models do not show the strong C features observed in superluminous
SNe~Ia, since the detonation leaves practically no unburned C in the
ejecta.

This conclusion is not sensitive to the (very uncertain) DDT criterion
we have used. In fact, even the prompt detonation model does not burn
significantly more material than the delayed detonation ones.
Due to the asymmetric propagation of the deflagration front and
the quick expansion in polar direction, the DDT will always occur close
to the poles and at a similar time after the start of the
deflagration, no matter what the transition criterion is.

With a \nni\ yield of 0.353\,\msol\ in the ejecta, the pure
deflagration model AWD3def is much fainter than the models involving
a detonation and clearly not a candidate for superluminous SNe~Ia. 
The main purpose for including this model in the present study was
to check the evolution of the thermonuclear burning if the DDT
fails. Nevertheless, it is interesting to look at the synthetic
observables of this model and discuss whether they may match any of
the observed peculiar SNe~Ia or other transients. Given that the
synthetic spectra of AWD3def resemble those of the pure deflagrations
in non-rotating WDs, which were found to agree fairly well with Type~Iax
SNe \citep{kromer2013a, fink2014a}, a potential connection to Type~Iax
SNe seems to be the obvious assumption. A detailed comparison to
the observed sample of SNe~Iax is beyond the scope of this paper, but
it is interesting to point out that rotation of the progenitor stars
seems to extend the possible outcomes of the pure deflagration
scenario. For example, our model AWD3def yields a \nni\ mass that is
very similar to the \nni\ yield of the non-rotating model N100def of
\citep{fink2014a}, while the total ejecta masses differ by about 30
percent (1.02 and 1.31\,\msol, respectively). This may help in
explaining some of the brighter members of the SN~Iax class
\citep[e.g.][]{magee2016a, barna2017a}.

\section{Conclusions}
\label{sec:conclusions}

The influence of rapid differential rotation of the progenitor WD on
the delayed-detonation scenario has been investigated.  To this end,
full-star hydrodynamic explosion simulations have been carried out in three
dimensions using three initial rotators that cover the expected range
of rotational energies.  A multispot ignition scenario was chosen for
the deflagration ignition and the DDT parametrization is based on a
critical Karlovitz number of $250$. We simulated the flame evolution
with the combustion-hydrodynamics code \textsc{leafs} and obtained
detailed nucleosynthesis yields and synthetic observables in
post-processing steps.

In the delayed-detonation models the deflagration quickly spreads
toward the poles, before it can significantly propagate in equatorial
direction. Thus, the DDT is always found to occur close to the poles
(in multiple spots) and when large fractions of the star are still
unburned.  The pre-expansion of the remaining fuel caused by the
deflagration is also found to be relatively limited.  Therefore, in
all models the detonation phase dominates the burning.  Consequently,
bright explosions ensue ($M_{\nni} > 1\,\msol$) that could be
potential candidates for superluminous SNe~Ia.  However, the
velocities of IMEs are significantly too high as compared to the low
line velocities seen in early-time spectra of observed superluminous
SNe~Ia.  Moreover, our synthetic spectra do not show the
characteristic carbon lines observed in superluminous SNe~Ia. We thus
conclude that the delayed-detonation scenario of rapidly rotating WDs
is incompatible with any of the observed SNe~Ia. Despite
the anisotropic flame evolution the final ejecta are on large scales
close to spherical and consequently observables do not show strong
viewing-angle sensitivity.

These results are robust with respect to several uncertainties
that are still present in the models: (i) details of the rotation laws
will not allow for much change in the explosion outcomes.  Within our
series of models that covers a broad range of rotational energies, the
predicted spectral signatures are very similar;  (ii) as the chosen
ignition conditions produce a strong deflagration phase in
non-rotating models, fewer ignition spots and, thus, weaker
deflagrations would lead to even stronger detonations;  (iii) the DDT
criterion is still uncertain.  However, in the models the dominant
effect is the anisotropy of the deflagration.  The flame will always
reach the low density edge of the WD in polar directions first and the
DDT has to occur close to the poles.  Due to the very slow propagation
of the deflagration in the equatorial plane, a somewhat later DDT will
not cause much more pre-expansion of unburned fuel there.

In agreement with \citet{pfannes2010b} we found that the problem of
ejecta velocities being too high is even more severe in the case of
prompt detonations.  This was demonstrated in one pure detonation
simulation (the AWD$3$ rotator) similar to one of the models of
\citet{pfannes2010b}.  There were some differences, though. Due to the
significantly lower NSE transition density of our improved detonation
scheme, much higher ejecta velocities and less IMEs than
in \citet{pfannes2010b} were found.  As the delayed detonations showed
too high IME velocities for the whole series of initial rotators, the
same trend will also hold for the corresponding prompt detonation
models. Thus, both, prompt detonations and the delayed detonations of
rapidly-rotating massive progenitors cannot explain the superluminous
SNe~Ia. Moreover, their synthetic lightcurves and spectra do not
resemble any other subclass of observed SNe~Ia.

We also investigated the case of a pure deflagration for the AWD$3$
rotator. Owing to the rapid rotation, turbulence is suppressed
perpendicular to the rotation axis, leading to an asymmetric evolution
of the deflagration flame and a low release of nuclear
energy. Consequently, the deflagration fails to fully unbind the
progenitor WD, even for our vigorous ignition scenario. Specifically,
we obtain an ejecta mass of $1.02\,\msol$ (of which $0.353\,\msol$ are
\nni), while $0.980\,\msol$ remain bound. The ejecta structure and
synthetic observables resemble those of SNe~Iax.

In summary, our results can be interpreted in two ways. The first
possibility is that no detonations occur in differentially rotating
super-Chandrasekhar mass WDs. In this case, deflagrations of
differentially rotating super-Chandrasekhar mass WDs could contribute
to the population of SNe~Iax. As an alternative explanation,
differentially rotating super-Chandrasekhar mass WDs may simply not
exist in nature. If, for instance, a mechanism would exist which
forces the WD into rigid rotation while accreting, other than those
modeled in the work of \citet{yoon2004a}, the maximum mass allowed
would be close to the canonical Chandrasekhar value.  In fact,
magnetic fields may provide such a transport mechanism as was
discussed by \citet{spruit1999a,spruit2004a} and more recently
investigated by means of numerical simulations by \citet{wei2015a}.
In principle, the argument is simple. If the WD has a poloidal
magnetic field $B_\text{p}$, differential rotation will wind-up this
field and will generate a toroidal component $B_{\Phi}$, increasing
linearly with time until the restoring force
($\propto B_\text{p}B_{\Phi}$) stops this process and reverts
it. These oscillations are damped by phase mixing \citep{spruit1999a}
and eventually will lead to rigid rotation.  However, whether or not
typical magnetic fields inside the progenitors of SNe~Ia are
sufficiently high that the damping time is short compared with the
accretion time has to be seen. Clearly, field strengths as found in
polars ($> 10^7$ G) would suffice \citep{spruit1999a,spruit2004a}.

\begin{acknowledgements}
  We gratefully acknowledge the Gauss Centre for Supercomputing (GCS)
  for providing computing time through the John von Neumann Institute
  for Computing (NIC) on the GCS share of the supercomputer JUQUEEN
  \citep{stephan2015a} at J\"ulich Supercomputing Centre (JSC). GCS is
  the alliance of the three national supercomputing centres HLRS
  (Universit\"at Stuttgart), JSC (Forschungszentrum J\"ulich), and LRZ
  (Bayerische Akademie der Wissenschaften), funded by the German
  Federal Ministry of Education and Research (BMBF) and the German
  State Ministries for Research of Baden-W\"urttemberg (MWK), Bayern
  (StMWFK) and Nordrhein-Westfalen (MIWF). MK, FR and RP acknowledge
  support from the Klaus Tschira Foundation. The research of FR was
  supported by the German Research Foundation (DFG) via the
  Collaborative Research Center SFB 881 ``The Milky Way System''. RP
  acknowledges support by the European Research Council under ERC-StG
  grant EXAGAL-308037. IRS was supported by the Australian Research
  Council Grant FT160100028. SAS acknowledges support from STFC
  through grant, ST/P000312/1. This work was supported by the Deutsche
  Forschungsgemeinschaft via the Transregional Collaborative Research
  Center TRR~33 ``The Dark Universe'' and by the DAAD/Go8
  German-Australian exchange program.
\end{acknowledgements}

\bibliographystyle{aa}

\end{document}